\documentclass[journal]{IEEEtran}
\usepackage{array}
\usepackage{url}
\usepackage{cite,graphicx,multirow,multicol,setspace,amsmath,wrapfig}
\usepackage{tabulary,amssymb,amsfonts,adjustbox,booktabs,qtree,tikz-qtree,synttree,rotating}
\usepackage[flushleft]{threeparttable}
\usepackage{makecell,eqnarray}
\usepackage{cuted}
\usepackage[linesnumbered,ruled]{algorithm2e}
\usepackage{algorithmic,textcomp,xcolor}
\usepackage{pifont}
\usepackage{commath}
\graphicspath{ {Images/} }

\usepackage{tikz}
\usetikzlibrary{shapes,arrows,decorations.markings,calc}
\usepackage[EULERGREEK]{sansmath}

\makeatletter
\let\MYcaption\@makecaption
\makeatother

\usepackage[font=footnotesize]{subcaption}

\makeatletter
\let\@makecaption\MYcaption
\makeatother

\usepackage{amsthm}
\theoremstyle{definition}

\makeatletter
\newcommand{\algorithmfootnote}[2][\footnotesize]{%
  \let\old@algocf@finish\@algocf@finish
  \def\@algocf@finish{\old@algocf@finish
    \leavevmode\rlap{\begin{minipage}{\linewidth}
    #1#2
    \end{minipage}}%
  }%
}

\makeatletter
\long\def\@makecaption#1#2{\ifx\@captype\@IEEEtablestring%
\footnotesize\begin{center}{\normalfont\footnotesize #1}\\
{\normalfont\footnotesize\scshape #2}\end{center}%
\@IEEEtablecaptionsepspace
\else
\@IEEEfigurecaptionsepspace
\setbox\@tempboxa\hbox{\normalfont\footnotesize {#1.}~~ #2}%
\ifdim \wd\@tempboxa >\hsize%
\setbox\@tempboxa\hbox{\normalfont\footnotesize {#1.}~~ }%
\parbox[t]{\hsize}{\normalfont\footnotesize \noindent\unhbox\@tempboxa#2}%
\else
\hbox to\hsize{\normalfont\footnotesize\hfil\box\@tempboxa\hfil}\fi\fi}
\makeatother

\def\myarm{1cm}
\def\myangle{0}
\tikzset{
  arm/.default=1cm,
  arm/.code={\def\myarm{#1}}, 
  angle/.default=0,
  angle/.code={\def\myangle{#1}} 
}

\tikzset{
    myncbar/.style = {to path={
        let
            \p1=($(\tikztotarget)+(\myangle:\myarm)$)
        in
            -- ++(\myangle:\myarm) coordinate (tmp)
            -- ($(\tikztotarget)!(tmp)!(\p1)$)
            -- (\tikztotarget)\tikztonodes
    }}
}

\begin{document}

\title{An Efficient Framework for Automated Screening of Clinically Significant Macular Edema}

\author{Renoh~Johnson~Chalakkal,~\IEEEmembership{Member,~IEEE,}
        Faizal~Hafiz,~\IEEEmembership{Member,~IEEE,}
        Waleed~H~Abdulla,~\IEEEmembership{Senior Member,~IEEE,}
        and~Akshya~Swain,~\IEEEmembership{Senior~Member,~IEEE}
\thanks{The authors are with School of Engineering, the University of Auckland, New Zealand. E-mail: rcha789@aucklanduni.ac.nz, faizalhafiz@ieee.org}%

\thanks{}}

\markboth{}
{}
%



\maketitle

\begin{abstract}

The present study proposes a new approach to automated screening of Clinically Significant Macular Edema (CSME) and addresses two major challenges associated with such screenings, \textit{i.e.}, \textit{exudate segmentation} and \textit{imbalanced datasets}. The proposed approach replaces the conventional exudate segmentation based feature extraction by combining a pre-trained deep neural network with meta-heuristic feature selection. A feature space over-sampling technique is being used to overcome the effects of skewed datasets and the screening is accomplished by a k-NN based classifier. The role of each data-processing step (\textit{e.g.}, \textit{class balancing, feature selection}) and the effects of limiting the region-of-interest to \textit{fovea} on the classification performance are critically analyzed. Finally, the selection and implication of operating point on Receiver Operating Characteristic curve are discussed. The results of this study convincingly demonstrate that by following these fundamental practices of machine learning, a basic k-NN based classifier could effectively accomplish the CSME screening.

\end{abstract}

\begin{IEEEkeywords}
Diabetic retinopathy, feature selection, fundus imaging, skewed datasets.
\end{IEEEkeywords}

%
\IEEEpeerreviewmaketitle





\section{Introduction}
\label{s:intro}

\IEEEPARstart{D}{iabetic} Macular Edema (DME) is the most common cause of vision loss in patients suffering from Diabetic Retinopathy (DR)~\cite{edtrs}. It is one of the important surreptitious manifestations of DR and is identified by the presence of retinal thickening within one Disc Diameters (1-DD) from the fovea center~\cite{edtrs}. The early treatment diabetic retinopathy study~\cite{edtrs} classifies DME into two distinct categories: non-Clinically Significant Macular Edema (Non-CSME) and Clinically Significant Macular Edema (CSME). In particular, the CSME variant of DME requires urgent clinical attention and it is therefore the focus of the current investigation. 

The CSME condition is often associated with the presence of hard exudates which are the lipo-protein deposits formed as a result of leakage from swellings of retinal vessels. Hence, CSME can be identified based on the position and number of hard exudates within 1-DD area around the fovea. Usually, the screening for CSME requires the evaluation of fundus photography and/or the optical coherence tomography imaging by an ophthalmologist/optometrist. This, perhaps, is the biggest challenge for the proliferation of early patient screening, especially, in the developing countries where the ratio of such specialists to patients is often quite low. Hence, there is a need for automated screening systems. To this end, several machine-learning based techniques have been developed for the screening of CSME directly from the fundus retinal images~\cite{histo1,histo2,back1,back2,Ravi2009,Osareh2009,RGB1,RGB2,Mo2018,Perdomo2016,Giancardo2011,sift,amfm}. Such approach is often quite cost-effective and does not require specialists.


Most of the existing automated screening approaches rely on segmentation of the exudates (\textit{hard and soft exudates}) from the area of interest (2-DD \textit{around the fovea}). The information about the presence, nature and number of the hard exudates at or near the fovea region is then used for screening purposes. Such approaches can be broadly classified into \textit{Local Schemes} and \textit{Global Schemes}.

The Local Schemes (LS) usually involves two steps. First, the hard-exudates in the retinal image are segmented. Next, the properties and nature of these exudates are quantified which form the basis for screening. In most of these schemes, the variation in the pixel intensity is used to segment the hard exudates, \textit{e.g.}, approaches based on \textit{image histogram}~\cite{histo1,histo2}, \textit{background pixel cancellation}~\cite{back1,back2}, \textit{edge detection}~\cite{Ravi2009,Osareh2009}, and \textit{color features}~\cite{RGB1,RGB2}. In addition, several different variants of LS based on \textit{deep learning} have recently been proposed in~\cite{Mo2018,Perdomo2016}. These variants essentially integrate both the steps of LS (\textit{i.e.}, segmentation and quantification of hard exudates) into the learning process. In contrast, the Global Schemes (GS) focus only on the presence of hard exudates. In particular, the detection of at-least one hard exudate implies the presence of DME in these schemes. To this end, several approaches have been developed, \textit{e.g.}, schemes based on \textit{thresholding and confidence map}~\cite{Giancardo2011}, \textit{visual word/group} using Scale Invariant Feature Transform (SIFT)~\cite{sift} and Amplitude Modulation-Frequency Modulation (AM-FM)~\cite{amfm} based features.

Although in the existing screening approaches, exudate segmentation has extensively been explored, the practical application of such approaches is often limited due to their sensitivity to possible image artifacts arising from several factors which include, but not limited to, poor illumination and calibration. Further, these approaches may also be affected by the variation in retinal image pigmentation across people of different age and ethnicity. However, the major limitation of this approach is neither the sensitivity to image artifacts nor the variation in retinal pigmentation; rather it is the need for manual identification of exudates within the retinal images. Such manual identification is often ambiguous, labor-intensive and time-consuming, which warrants the need for an alternative to exudate segmentation.

\begin{figure*}[!t]
\centering
\begin{adjustbox}{max width=0.98\textwidth}
\tikzstyle{block} = [rectangle, draw, fill=white, 
    text width=5em, text centered, rounded corners, minimum height=4em,line width=0.02cm]
\tikzstyle{block1} = [rectangle, draw, fill=white, 
    text width=6.5em, text centered, rounded corners, minimum height=4em,line width=0.02cm]
\tikzstyle{line} = [draw, -latex']
\tikzstyle{cloud} = [draw, ellipse,fill=white, node distance=4.8cm,
    minimum height=3em, line width=0.02cm]

\begin{tikzpicture}[auto]
    \centering
    \node [block,fill=blue!20] (images) {Retinal Images};
    \node [block1, right of=images, node distance = 3cm,fill=orange!20] (prep) {Optional Fovea Segmentation};
    \node [block, right of=prep, node distance = 3cm,fill=orange!20] (fe) {Feature Extraction};
    \node [block, right of=fe, node distance = 3cm,fill=orange!20] (cb) {Class Balancing};
    \node [block, right of=cb, node distance = 3cm,fill=orange!20] (fs) {Feature Selection};
    \node [block, right of=fs, node distance = 3cm,fill=green!20] (knn) {k-NN};
    
    \path [line,ultra thick] (images) -- (prep);
    \path[line,ultra thick] (prep)--(fe);
    \path[line,ultra thick] (fe)--(cb);
    \path[line,ultra thick] (cb)--(fs);
    \path[line,ultra thick] (fs)--(knn);
    \path[line,ultra thick] ([yshift=-10pt]knn.east) -- node[align=center, below] {Non-CSME} +(2cm,0pt);
    \path[line,ultra thick] ([yshift=10pt]knn.east) -- node[align=center, above] {CSME} +(2cm,0pt);
\end{tikzpicture}  
\end{adjustbox}
\caption{The proposed framework for automated screening of CSME.}
\label{f:DRBlock}
\end{figure*}
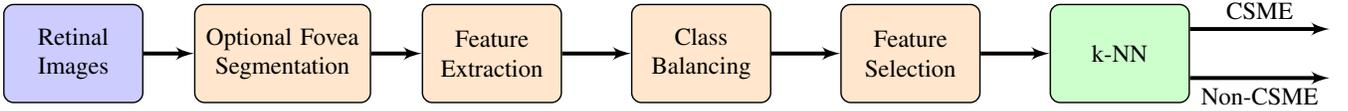

The deep learning based approach offers a possible alternative to exudate segmentation, as demonstrated by various researchers in bio-medical imaging. For instance, Deep Convolution Neural Network (DCNN) has been used to grade the severity of DR directly from the fundus images~\cite{Pratt2016,Quellec2017,Gargeya:Leng:2017}. However, the extension of the similar approach to CSME screening is not quite straight forward and poses several challenges, \textit{e.g.}, the need for large number of \textit{labeled} fundus images and increased complexity. Nevertheless, while the use of DCNN as a complete screening tool may be challenging, it can effectively be used in limited capacity as a \textit{feature extractor} to replace exudate segmentation. In particular, the outputs from the last fully connected layer of DCNN can be used as \textit{features} which often outperforms the conventional \textit{hand-crafted} features~\cite{Xu:Mo:2014}. 

Pursuing to these arguments, it is easy to follow that a \textit{pre-trained} DCNN may be considered as the \textit{feature extractor}. A caveat to this remedy is the possible presence of several \textit{redundant} and/or \textit{irrelevant} features, as it is likely that the network is originally trained for a different, and often non-medical, application, \textit{e.g.}, AlexNet, GoogLeNet~\cite{alex, google}. Such irrelevant features can have detrimental effects on the generalization capabilities of the classifier. The identification of parsimonious subset of \textit{relevant} features through feature selection is, therefore, crucial additional step when a \textit{pre-trained} network is used as the feature extractor.   

Further, the class distribution in most of the fundus image datasets is often heavily skewed, \textit{i.e.}, the number of CSME images is significantly lower. For instance, the ratio of Non-CSME to CSME images is approximately 3 in several well-known benchmark datasets such as MESSIDOR~\cite{Messidor2014}, IDRiD~\cite{Idrid} and UoA-DR~\cite{Chalakkal2017a}, which indicates a significant class imbalance~\cite{Das:Datta:2018}. While such skewed datasets are common occurrence in most of the medical image analysis problems~\cite{Lerner2007, Haixiang2017}, it represents a significant challenge to induce a balanced learning hypothesis. Such class imbalance coupled with small data size can have detrimental effects on the generalization capabilities of the classifier~\cite{Chawla2002,Das:Datta:2018}. However, the effects of class imbalance are often neglected in most of the existing screening approaches~\cite{Mo2018,Perdomo2016}. 

To summarize, skewed databases and dependency on exudate segmentation are some of the major factors which limit the proliferation of automated screening of CSME. 

To bridge this gap, the present study proposes a simple screening approach which replaces \textit{exudate segmentation} and addresses the issue of \textit{imbalanced} datasets. In particular, it is shown that by following few fundamental \textit{good} practices of machine learning, this task can be accomplished by a simple \textit{k}-NN based classifier. Such an approach is suitable for large scale deployment owing to its simplicity and cost-effectiveness. The proposed screening approach uses a pre-trained DCNN in a limited role as a \textit{feature extractor} to replace exudate segmentation. The efficacy of the extracted features is further determined through a meta-heuristic feature selection approach. Next, the issue of \textit{class imbalance} is addressed by a feature space over-sampling technique. Finally, a simple binary classifier is induced by k-NN for CSME screening purposes. 

The key contributions of this investigation are:
\begin{itemize}
    \item A combination of pre-trained DCNN and meta-heuristic feature selection is proposed to replace conventional exudate segmentation. A detailed comparative analysis is carried out considering 11 pre-trained DCNNs and 2 meta-heuristic feature selection algorithms.
    
    \item The role of \textit{class balancing} is critically analyzed considering a simple feature space up-sampling technique. The degree of oversampling and its effects on the classifier performance are also analyzed.
    
    \item The fovea segmentation (1-DD around the fovea) is considered as an optional image pre-processing step and its effects on the CSME screening is investigated.

    \item The selection of the ROC operating point and its implications on related screening costs are analyzed.
    \item The efficacy of the proposed approach has been evaluated considering 3 well-known public domain retinal image databases.
    
\end{itemize}     

The rest of the article is organized as follows: A brief description of the databases used in this study is given in section \ref{s:database}. The optional fovea segmentation stage is detailed in section \ref{s:fovseg}. Section \ref{s:feat_ex} introduces the DCNN--based feature extraction technique. The method used for negotiating the class imbalance problem is discussed in section \ref{s:cls_imb}. Feature selection technique used for reducing the feature set dimension is discussed in section \ref{s:fet_sel}. Section \ref{s:results} discuss the results obtained by each step in detail. Finally, the paper concludes in section \ref{s:conclusion} by providing possible future direction of the research.

\section{Investigation Framework}
\label{s:invframe}

The goal of this study is to develop a simplified approach to automate CSME screening. In particular, the objective here is to address the issues associated with exudate segmentation and skewed datasets. To this end, this study proposes a simple framework which involves few data processing steps as outlined in Fig.~\ref{f:DRBlock}. The proposed framework relies on a combination of pre-trained DCNN and a meta-heuristic feature selection to replace conventional {exudate segmentation} based feature extraction. The effects of skewed datasets are overcome by the introduction of synthetic samples in the feature space. Further, it is expected that limiting the field of view of retinal images to 1-DD around fovea may improve the screening accuracy. To investigate this further, the fovea segmentation is also considered as an optional data-processing step. In the following, these steps are discussed in detail.

\begin{table}[!t]
  \small
  \centering
  \caption{Retinal Image Databases}
  \label{tab:1}%
  \begin{adjustbox}{width=0.47\textwidth}
  \begin{threeparttable}
    \begin{tabular}{ccccc}
    \toprule
    \textbf{Database} & \textbf{Remark} & \makecell{\textbf{Assigned} \\ \textbf{Class}} & \makecell{\textbf{Number of} \\ \textbf{Images}} & $^\dagger$\textbf{Ratio} (\boldmath$\%$) \\[0.5ex]
    \midrule
    \multirow{4}{*}{MESSIDOR} & \makecell{Grade 0} & Non-CSME & 964   & 81.2 \\[0.5ex]
          & \makecell{Grade 1} & Non-CSME & 73    & 6.1 \\[0.5ex]
          & \makecell{Grade 2} & CSME  & 150   & 12.6 \\[0.5ex]
    \cmidrule{2-5}          & Total  & -     & 1187  & - \\[0.5ex]
    \midrule
      \multirow{4}{*}{UoA-DR} & Healthy & Non-CSME & 56   & 28 \\[0.5ex]
          & DR only & Non-CSME & 70    & 35 \\[0.5ex]
          & CSME & CSME  & 74   & 37 \\[0.5ex]
    \cmidrule{2-5}          & Total  & -     & 200  & - \\[0.5ex]
    \midrule
     \multirow{4}{*}{IDRiD} & Healthy & Non-CSME & 168   & 32.5 \\[0.5ex]
          & DR only & Non-CSME & 54    & 10.5 \\[0.5ex]
          & DME 1 & Non-CSME  & 51   & 9.9 \\[0.5ex]
          & DME 2 & CSME  &  243  & 47.1 \\[0.5ex]
    \cmidrule{2-5}          & Total  & -     & 516  & - \\
    \bottomrule
    \end{tabular}%
    \begin{tablenotes}
        \item $^\dagger$ ratio of images in a particular class to the total number of images
    \end{tablenotes}
    \end{threeparttable}
\end{adjustbox}
\end{table}%

\subsection{Database}
\label{s:database}

In this study, the following three retinal image databases are considered: MESSIDOR~\cite{Messidor2014}, UoA-DR~\cite{Chalakkal2017a} and IDRiD~\cite{Idrid}. These databases are combined to give a total of 1903 retinal images as shown in Table~\ref{tab:1}. Further, since the focus of this study is only on clinically significant DME (CSME), the other grades of DME are grouped into one class and is referred here as `Non-CSME'. For example, in the MESSIDOR database, only the grade-2 DME images are labeled as CSME images, and the remaining grades of DME are labeled as Non-CSME. 

It is worth emphasizing that the CSME images form only a small portion of each dataset. For instance, in the MESSIDOR database, only $12\%$ of images belong to CSME class. Similarly, in UoA-DR and IDRiD database, the number of CSME images are 74 ($37\%$ of the total images)  and 243 ($47\%$ of the total images), respectively. Hence, there exists a significant \textit{class imbalance} in the considered datasets. This issue will be discussed at length in Section~\ref{s:cls_imb}.

Further, each image is analyzed following the image quality analysis approach in~\cite{Chalakkal2019a} which ensures that the image under consideration is of acceptable quality according to the medically suitable retinal image standard~\cite{abdel}. Following this step, a total of 1767 images are determined to be of acceptable quality. These images are randomly split into training (80\% images) and testing (20\% images) datasets following the principle of cross-validation. The training dataset consists of 1399 images (346 CSME and 1053 Non-CSME) and the testing set consists of 368 images (94 CSME and 274 Non-CSME).

\begin{figure}[!t]
\centering
    \includegraphics[width=0.4\textwidth]{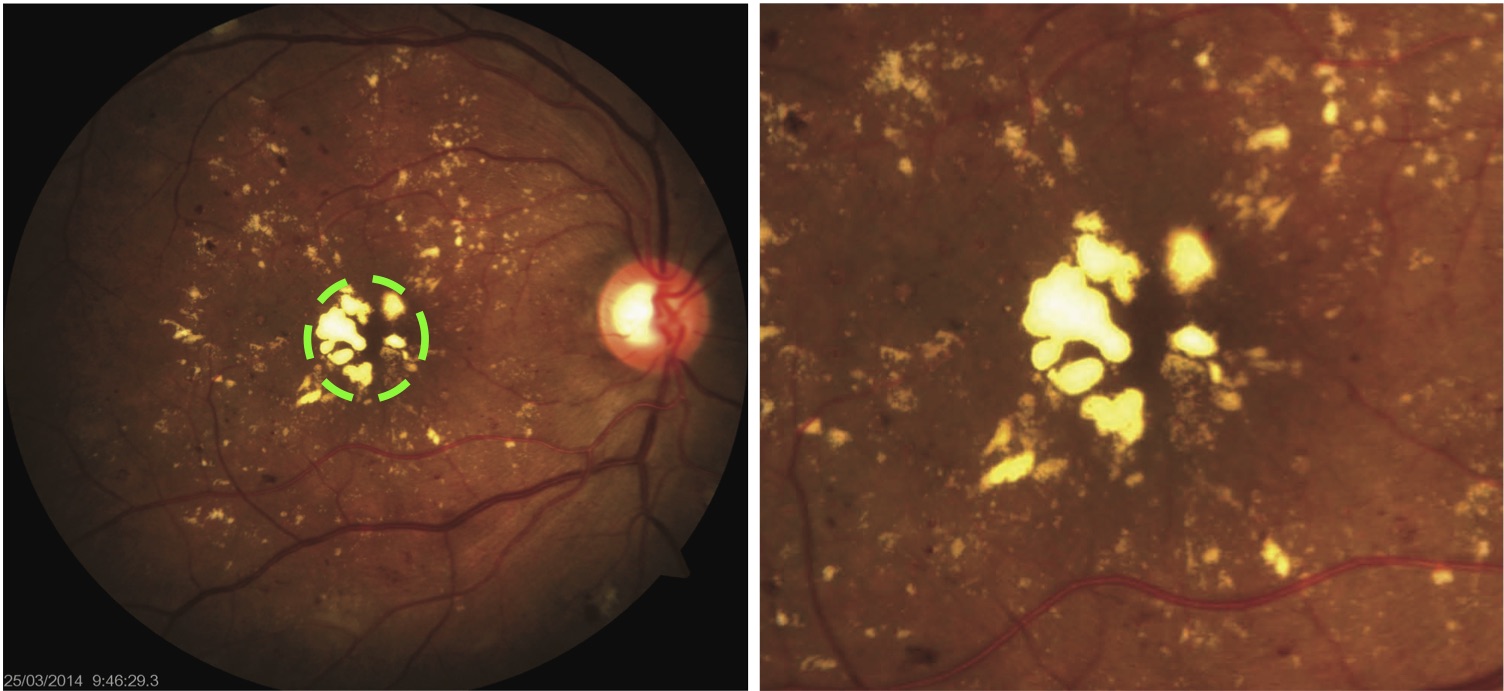}
    \caption{The effects of fovea segmentation. The left and right image respectively represent a full retinal image and the corresponding segmented fovea.}
    \label{ch6_fig3}
\end{figure}
\subsection{Fovea Segmentation}
\label{s:fovseg}

In practice, it is observed that limiting the field of view to the Region-Of-Interest (ROI), often reduces the computational complexity without any significant trade-off in the classification performance~\cite{Chalakkal2018a}. To investigate this further, in this study, `\textit{fovea-segmentation}' is considered as an optional pre-processing stage, where 1 Disc Diameter (DD) around fovea is selected as the Region-Of-Interest (ROI). To this end, the fovea detection technique outlined in~\cite{Chalakkal2018a} is followed to extract the ROI. Fig. \ref{ch6_fig3} shows the original and segmented image obtained following this approach. 

For the remainder of this study, the retinal images with and without segmentation are referred to as `\textit{fovea}' and `\textit{full}' images, respectively. The effects of fovea segmentation are discussed at length in Section \ref{res:fov_seg}. 

\subsection{Feature Extraction}
\label{s:feat_ex}

Since feature extraction is the first key step in any machine learning algorithm, it is appropriate to briefly discuss some of the existing feature extraction approaches for sake of completeness. Most of the existing approaches~\cite{Joshi_2018}, rely on the \textit{a priori} expert knowledge to extract the relevant structures, \textit{e.g.}, hard exudates, micro-aneurysms. The information about these structures is subsequently used to extract features using distinct techniques such as Color Histogram, Histogram of Oriented Gradient, Morphological Operations, Scale-Invariant Feature Transform, Local Binary Pattern, Auto-Encoders, restricted Boltzmann Machines, Sparse Representations and Wavelet Transforms~\cite{Joshi_2018}. One of the major limitations of these approaches is the need for the \textit{a priori} expert knowledge. Such `hand-crafted' features are often application-specific and can not easily be generalized for other clinical settings. 

In few recent studies, pre-trained Deep Convolution Neural Networks (DCNN) have been used for analysis of medical images~\cite{Chen_2015,Gustavo_2015}. The DCNN have been shown to be effective, even-though these are trained on generic non-medical datasets. Further, since the feature extraction process is embedded in DCNN, the pre-trained DCNN can be used to extract features from the retinal images. This study, therefore, considers pre-trained DCNN as the feature extractor where the outputs of the last fully connected layer are used as the features. This step reduces the complexity of the feature extraction process by eliminating the need for \textit{a priori} expert knowledge. 

In this study, 11 pre-trained DCNNs are being considered and a  detailed comparative evaluation of these DCNNs, as the feature extractor, is given in Section~\ref{s:resDCNN}. These DCNNs have been pre-trained on ImageNet database~\cite{imagenet}.   

It is worth noting that the role of pre-trained DCNN is limited only to the \textit{feature extraction}. During the preliminary phases of this investigation, the performance of DCNN as the classifier was found to be unsatisfactory. Therefore, the research is not pursued in this direction.
\subsection{Class Imbalance}
\label{s:cls_imb}

The skewed datasets can be balanced either by generating synthetic samples from the minority class or by removing samples from the majority class. It has been shown that by removing majority samples arbitrarily (\textit{say using the random under-sampling}) may filter important data~\cite{Lerner2007}. This study, therefore, focuses on the over-sampling of minority class. The required synthetic samples can be generated either in the \textit{image space} or in the \textit{feature space}. In the context of class imbalance problem, CSME and Non-CSME classes are respectively referred here as the `\textit{minority}' and the `\textit{majority}' class. 

In the \textit{image space}, the synthetic samples can be generated using affine transformations such as scaling, rotation, random cropping, color jittering, Gaussian blur, shears and filtering~\cite{Hussain2017}. While such approaches are effective, they are often application-specific and computationally involved. In contrast, the complexity of oversampling can be reduced in the \textit{feature space} as the synthetic samples can  directly be generated by few numerical operations on the existing features. Further, such approach is application independent; since the oversampling is carried out in the feature space. Hence, in this study, a simple feature space oversampling strategy known as Synthetic Minority Over-sampling Technique (SMOTE)\cite{Chawla2002}, is being considered.

In SMOTE, a synthetic sample is generated in the geometric space between each minority sample and the corresponding $k^{th}$ nearest minority neighbor. The number of synthetic samples is controlled by the user specified over-sampling ratio ($r$), which is given by,
\begin{align}
    \label{eq:ratio}
    r=\frac{\text{Total Number of Synthetic Samples}}{\text{Total Number of Original Minority Class Samples}}
\end{align}
It is worth noting that while the higher value of $r$ may improve the \textit{sensitivity} (accuracy of the minority class), such improvement often involves a trade-off in the specificity (accuracy of the majority class). A proper selection of the over-sampling ratio is therefore critical to balance the overall performance of the classifier. A detailed investigation on choice of $r$ and its impact on the classification performance is discussed in Section~\ref{s:resClassBalance}.



\subsection{Feature Selection}
\label{s:fet_sel}

In this study, several pre-trained DCNN are being considered for feature extraction purposes. While this eliminates the need for \textit{exudate segmentation}, it is likely that some of the extracted features are irrelevant/redundant as the considered DCNN have been trained to classify generic, non-medical, images. The selection of relevant features is therefore a crucial processing step and it is one of the fundamental problem of machine learning. To understand this problem, let `$\mathcal{X}_{prime}$' denote a set of $n$-features extracted by a pre-trained DCNN as follows:
\begin{align}
    \mathcal{X}_{prime} = \begin{bmatrix} x_1 & x_2 & \dots & x_n \end{bmatrix}
\end{align}
The goal of feature selection is to identify the optimum subset of features denoted by `$\mathcal{X}^\star$' as follows:
\begin{align}
    \label{eq:fsp}
    \mathcal{X}^{\star} & = \Big\{ \mathcal{X} \subset \mathcal{X}_{prime} \ | \    J(\mathcal{X}) = \min \limits_{\forall{\mathcal{X}_i} \subset \mathcal{X}_{prime}} J(\mathcal{X}_i) \Big\}
\end{align}
where, `$J(\cdotp)$' denotes a suitable criterion function.

It is worth noting that the search for the optimal feature subset involves the examination of all possible subset combinations due to correlation among features~\cite{Guyon:Isabelle:2003,Xue:Zhang:2016,Hafiz:Swain:2017,Hafiz:Swain:2019,Hafiz:Swain:2018}, \textit{i.e.}, for the given $n$ number of features, the search space is composed of $2^n$ possible subsets. The exhaustive search over all possible subset is therefore not feasible even for moderate number of features. The feature selection problem is known to be NP-Hard~\cite{Guyon:Isabelle:2003} and therefore requires efficient search strategies. To this end, several feature selection approaches have been developed over the past seven decades. Most of the existing approaches can be categorized based on the criterion function (\textit{e.g., filters vs. wrappers}) or the search strategy (\textit{e.g., deterministic vs. meta-heuristic}). A detailed treatment on these approaches can be found elsewhere in~\cite{Guyon:Isabelle:2003,Xue:Zhang:2016}. In this study, a \textit{meta-heuristic wrapper} approach is followed. The rationale behind this decision is two-fold:
\begin{itemize}
    \item Most of the existing deterministic approaches require the selection of \textit{subset size}, which is usually not known \textit{a priori}. Hence, an additional auxiliary routine is often required to estimate the optimum subset size. This limitation can be overcome by most of the meta-heuristic approaches~\cite{Hafiz:Swain:2018}.
    
    \item For the given classifier, wrappers are relatively more accurate than filters; albeit at the expense of increased complexity~\cite{Guyon:Isabelle:2003}. Hence, the selection of a particular approach essentially involves a compromise in either precision or complexity; it is, therefore, pre-dominantly dependent on the number of features, `$n$'. In this study, the number of features is limited to 1000 (\textit{i.e.}, $n=1000$, discussed in Section~\ref{s:resDCNN}), hence, the wrapper approach is more appropriate.  
\end{itemize}  

Pursuing to these arguments, two classical meta-heuristic search algorithms, Genetic Algorithm (GA)~\cite{Siedlecki1989} and Binary Particle Swarm Optimization (BPSO)~\cite{Kennedy1997}, are being applied as a wrapper to \textit{k}-NN classifier. Each search agent (\textit{e.g.}, a \textit{chromosome} in GA or a \textit{particle} in BPSO) encodes a \textit{feature subset} in an $n$ - dimensional binary vector. For example, the $i^{th}$ search agent is denoted by `$\beta_i$' and it is represented as:
\begin{align}
    \beta_i & = \begin{bmatrix} \beta_{i,1} & \beta_{i,2} & \dots & \beta_{i,n} \end{bmatrix}, \\
    \text{where, } & \beta_{i,m} \in \{0,1\}, \quad m=1,2,\dots n \nonumber
\end{align}
The $m^{th}$ feature ($x_m \in \mathcal{X}_{prime}$) is included into the candidate subset provided the corresponding bit in the particle, `$\beta_{i,m}$' is set to `$1$'.

Further, each feature subset under consideration is evaluated based on classification performance of \textit{k}-NN over 10-fold stratified cross-validation as follows:
\begin{align}
    \label{eq:fit}
    J(\mathcal{X}_i) = 1 - \sum \limits_{k=1}^{10} AUC_{i,k}
\end{align}
where, `$\mathcal{X}_i$' and `$J(\mathcal{X}_i)$' denote respectively the feature subset under consideration and the corresponding criterion function; `$AUC$' denotes the Area Under the ROC Curve; and `$AUC_{i,k}$' denotes the AUC obtained with $\mathcal{X}_i$ over the $k^{th}$-fold of the training data. This procedure is outlined in Algorithm~\ref{al:crf}.

\begin{algorithm}[!t]
    \small
    \SetKwInOut{Input}{Input}
    \SetKwInOut{Output}{Output}
    \SetKwComment{Comment}{*/ \ \ \ }{}
    \Input{Search Agent, $\beta_i$}
    \Output{Criterion Function, $J(\mathcal{X}_i)$}
    \BlankLine
    Set the $i^{th}$ feature subset to null vector, \textit{i.e.}, $\mathcal{X}_i=\varnothing$\\
    \BlankLine
    \Comment*[h] {Decode the search agent}\\\nllabel{line:crf1}
    \For{m = 1 to $n$} 
        { \BlankLine
           \If{$\beta_{i,m}=1$}
           {\BlankLine
             $\mathcal{X}_i \leftarrow \{ \mathcal{X}_i \cup x_m \}$  \Comment*[h] {add the $m^{th}$ feature}\\
            \BlankLine
           }
          \BlankLine
        } 
    \BlankLine\nllabel{line:crf2}
    \BlankLine
    \Comment*[h] {10-fold stratified cross-validation}\\
    \BlankLine
    \For{k = 1 to 10}
     { \BlankLine
        Determine `$AUC$' corresponding to $\mathcal{X}_i$ and $k^{th}$-fold of the training data, see~\cite{Fawcett:ROC:2006}
      }
    \BlankLine    
    \Comment*[h] {Evaluate the Criterion Function}\\
    \BlankLine
    $J(\mathcal{X}_i) = 1 - \sum \limits_{k=1}^{10} AUC_{i,k}$
    \BlankLine
\caption{Evaluation of Criterion Function, $J(\cdotp)$}
\label{al:crf}
\end{algorithm}

It is easy to follow that, in this study, the feature selection is approached as an optimization problem. The objective here is to minimize $J(\cdotp)$, given by (\ref{eq:fit}), to identify the optimum feature subset, given by (\ref{eq:fsp}). To this end, the classical variant of GA and BPSO are being considered and the corresponding parameters are set as follows:
\begin{itemize}
    \item GA: \textit{binary tournament} selection; \textit{parameterized-uniform} crossover with probability $p_c=0.8$; \textit{flip-bit} mutation operator with probability $p_m=\displaystyle \frac{1}{n}$; population size $30$.
    \item BPSO: \textit{inertia weight}, $\omega=1$; \textit{acceleration constants} $[c_1,c_2]=[2,2]$; population size $30$.
\end{itemize}

To account for the inherent stochastic nature of the algorithms, `$R$' independent runs of each algorithm are executed ($R=40$), as outlined in Line \ref{Ch6_line:1}-\ref{Ch6_line:2} (Algorithm~\ref{al:ovfs}). Each run is set to terminate after $6000$ Function Evaluations (FEs). The `\textit{best}' subset, out of $R$ runs of each algorithm, is denoted by `$\mathcal{X}^*$' and it is considered for further analysis, as outlined in Line \ref{Ch6_line:3}-\ref{Ch6_line:4} (Algorithm~\ref{al:ovfs}). The outcomes of feature selection are discussed in detail in Section~\ref{res:fet_sel}.

\begin{algorithm}[!t]
    \small
    \SetKwInOut{Input}{Input}
    \SetKwInOut{Output}{Output}
    \SetKwComment{Comment}{*/ \ \ \ }{}
    \Input{Full feature set, $\mathcal{X}_{prime}$ and Class labels, $\mathcal{Y}$}
    \Output{Identified Feature Subset, $\mathcal{X}^*$}
    \BlankLine
    \BlankLine
    \Comment*[h] {Search for the optimal feature subset}\\
    \BlankLine
    Select a search algorithm (GA and BPSO in this study)\\
    Perform $R$ independent runs of the selected search algorithm \nllabel{Ch6_line:1}\\
    $\Omega \leftarrow \varnothing$\\
    \BlankLine
    \For{k = 1 to $R$} 
        { \BlankLine
           Identify the feature subset ($\mathcal{X}_{run}^k$) with the minimum $J(\cdotp)$\\
          \BlankLine
          $\Omega \leftarrow \{ \Omega \cup \mathcal{X}_{run}^k \} $\\
          \BlankLine
        } 
     \BlankLine \nllabel{Ch6_line:2}
     \Comment*[h] {Select feature subset out of $R$-runs}\\
     \BlankLine
     Select the feature subset ($\mathcal{X}^*$) with the minimum $J(\cdotp)$, \textit{i.e.},\nllabel{Ch6_line:3}
     \BlankLine
     $\mathcal{X}^* = \{ \mathcal{X}_{run}^m | J(\mathcal{X}_{run}^m) = \arg \min \limits_{k=1:R} J(\mathcal{X}_{run}^k), \ \  \forall \ \mathcal{X}_{run}^k \in \Omega \}$\nllabel{Ch6_line:4}\\
     \BlankLine
\caption{Meta-heuristic Feature Selection}
\label{al:ovfs}
\end{algorithm}

\section{Results and Discussion}
\label{s:results}

The accurate screening of CSME is dependent on several factors. The goal of this study is therefore to investigate whether few fundamental data processing tasks (\textit{e.g.}, feature selection and class balancing) can judiciously be used to achieve better performance by using a simple classifier such as k-NN. In the following subsections, the role of each pre-processing step is critically analyzed and discussed. 


First, a detailed comparative analysis of 11 pre-trained DCNNs, as an alternative to exudate segmentation, is discussed in Section~\ref{s:resDCNN}. Next, the role of class balancing is examined by considering several degrees of oversampling ratio in Section~\ref{s:resClassBalance}. This is followed by identification of relevant subset of features, which is discussed in Section~\ref{res:fet_sel}. The selection of operating point on the ROC curve and the associated cost-benefit analysis are discussed in Section~\ref{res:roc_opera}. Finally, the effects of optional fovea segmentation are discussed in Section~\ref{res:fov_seg}.

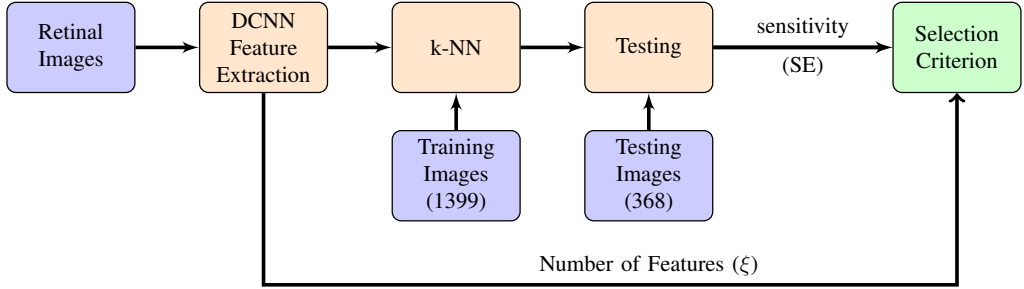
\begin{figure*}[!t]
\centering
\begin{adjustbox}{max width=0.75\textwidth}
\tikzstyle{block} = [rectangle, draw, fill=white, 
    text width=5em, text centered, rounded corners, minimum height=4em,line width=0.02cm]
\tikzstyle{block1} = [rectangle, draw, fill=white, 
    text width=6em, text centered, rounded corners, minimum height=4em,line width=0.0cm]
\tikzstyle{line} = [draw, -latex']
\tikzstyle{cloud} = [draw, ellipse,fill=white, node distance=4.8cm,
    minimum height=3em, line width=0.02cm]

\begin{tikzpicture}[auto]
    \centering
    \node [block,fill=blue!20] (images) {Retinal Images};
    \node [block, right of=images, node distance = 3cm,fill=orange!20] (fe) {DCNN Feature Extraction};
    \node [block, right of=fe, node distance = 3cm,fill=orange!20] (kt) {k-NN};
    \node [block, right of=kt, node distance = 3cm,fill=orange!20] (val) {Testing};
    \node [block, below of=kt, node distance = 2cm,fill=blue!20] (ts) {Training Images (1399)};
    \node [block, below of=val, node distance = 2cm,fill=blue!20] (tes) {Testing Images (368)};
    \node [block, right of=val, node distance = 4.8cm,fill=green!20] (sel) {Selection Criterion};
    \node [below of=tes, node distance = 1.4cm] (xi) {Number of Features ($\xi$)};

    \path [line,ultra thick] (images) -- (fe);
    \path[line,ultra thick] (fe)--(kt);
    \path[line,ultra thick] (ts)--(kt);
    \path[line,ultra thick] (kt)--(val);
    \path[line,ultra thick] (tes)--(val);
    \path[line,ultra thick] (val)--(sel);
     \path [line, ultra thick] (val) --  node [align=center, above]{sensitivity}node [align=center, below]{(SE)} (sel); 
    \draw[->,ultra thick] (fe)  to [myncbar,angle=270,arm=3.7]   (sel);
\end{tikzpicture}  
\end{adjustbox}
\caption{Selection of DCNN-based feature extractor.}
\label{f:DCNNSel}
\end{figure*}
\begin{table*}[!t]
  \centering
  \small
  \caption{Comparative Evaluation of DCNN based Feature Extractors}
  \label{t:compDCNN}%
  \begin{adjustbox}{max width=0.95\textwidth}
    \begin{tabular}{cccccccc}
    \toprule
    \multirow{2}[2]{*}{\textbf{DCNN}} & \multirow{2}[2]{*}{\makecell{\textbf{Number of}\\ \textbf{Features (\boldmath$\xi$)}}} & \multicolumn{3}{c}{\textbf{Full Image}} & \multicolumn{3}{c}{\textbf{Fovea}} \\[0.5ex]
    \cmidrule{3-8}          &       & {\makecell{\textbf{Sensitivity} \\ \textbf{(SE)}}} & {\makecell{\textbf{Specificity} \\ \textbf{(SP)}}} & \makecell{\textbf{Overall}\\ \textbf{Accuracy (\boldmath$\eta$)}} & {\makecell{\textbf{Sensitivity} \\ \textbf{(SE)}}} & {\makecell{\textbf{Specificity} \\ \textbf{(SP)}}} & \makecell{\textbf{Overall}\\ \textbf{Accuracy (\boldmath$\eta$)}} \\[0.5ex]
    \midrule
    AlexNet~\cite{alex} & 4096  & 0.6452 & 0.9418 & 0.8668 & 0.6809 & 0.9453 & 0.8777 \\[0.5ex]
     GoogLeNet~\cite{google} & 1024  & 0.5806 & 0.9382 & 0.8478 & 0.7021 & 0.9489 & 0.8859 \\[0.5ex]
    Inceptionv3~\cite{inceptionv3} & 1000  & 0.5699 & 0.9309 & 0.8397 & 0.7021 & 0.938 & 0.8777 \\[0.5ex]
    \textbf{InceptionResnet-v2}~\cite{inceptionresnetv2} & \textbf{1000} & 0.6237 & 0.9309 & 0.8533 & \textbf{0.8085} & {0.9526} & {0.9158} \\[0.5ex]
    ResNet18~\cite{resnet50} & 512   & 0.5914 & 0.9455 & 0.856 & 0.7021 & 0.9708 & 0.9022 \\[0.5ex]
    \textbf{ResNet50}~\cite{resnet50} & \textbf{1000} & \textbf{0.6452} & {0.9309} & {0.8587} & 0.7553 & 0.9416 & 0.894 \\[0.5ex]
    ResNet101~\cite{resnet50} & 1000  & 0.6129 & 0.9273 & 0.8478 & 0.7447 & 0.9672 & 0.9103 \\[0.5ex]
    VGG16~\cite{vgg16} & 4096  & 0.5914 & 0.9236 & 0.8397 & 0.6596 & 0.9526 & 0.8777 \\[0.5ex]
    VGG19~\cite{vgg16} & 1000  & 0.5806 & 0.9273 & 0.8397 & 0.7128 & 0.9234 & 0.8696 \\[0.5ex]
    DenseNet201~\cite{densenet} & 1000  & 0.5914 & 0.92  & 0.837 & 0.6809 & 0.9672 & 0.894 \\[0.5ex]
    SqueezeNet~\cite{squeezenet} & 1000  & 0.6022 & 0.9382 & 0.8533 & 0.5745 & 0.9453 & 0.8505 \\
    \bottomrule
    \end{tabular}%
  \end{adjustbox}
\end{table*}%

\subsection{Comparative evaluation of DCNN--based feature extraction}
\label{s:resDCNN}

In this study, a total of 11 pre-trained DCNNs are being considered for the 
purpose of feature extraction. Fig.~\ref{f:DCNNSel} shows overall procedure followed for the comparative evaluation of these DCNNs. The objective here is to select a DCNN with a better trade-off in sensitivity and number of features ($\xi$), as highlighted in Fig.~\ref{f:DCNNSel}. To this end, the extracted features from each DCNN are used to induce a classifier using 3-NN and the training images (see Section~\ref{s:feat_ex}). Subsequently, the classification performance of these trained classifiers over the testing images (see Section~\ref{s:feat_ex}) is determined. 

It is worth noting that, in practice, the value of `$k$' can be tuned for a given dataset to further optimize the performance. This approach, however, could not be followed in this study as the dataset considered in this study are not fixed, given that several distinct degree of over-sampling ratio are being considered. Further, earlier investigation suggest that lower values of $k$ are usually appropriate for medical images~\cite{Hand2003}.

Table~\ref{t:compDCNN} shows the results of the comparative evaluation on both full and fovea images. The results for \textit{fovea} images clearly indicate that `InceptionResnet-v2' yields the highest sensitivity with relatively fewer features. It is therefore selected as the feature extractor for the \textit{fovea} images. Further, for the \textit{full} images, both `AlexNet' and `ResNet50' can be considered for feature extraction purposes as shown in Table~\ref{t:compDCNN}. However, in this study , `ResNet50' is selected since it yields sensitivity similar to `AlexNet' but with significantly lower number of features.

\subsection{Role of Class Balancing}
\label{s:resClassBalance}

Given that the retinal image database being used is heavily skewed, it should be balanced before induction of any classifier. Therefore, the minority/positive class samples are over-sampled in feature space using SMOTE (as discussed in Section~\ref{s:cls_imb}). The first step of this process is to determine the appropriate over-sampling ratio $r$, which is given by (\ref{eq:ratio}). To determine the effects of over-sampling on the classification performance, several degrees of oversampling ratios are considered with $r\in[0,2]$. Note that $r=0$ denotes the original dataset without oversampling. For each degree of oversampling, the classification performance of 3-NN over the test images is determined. 
\begin{figure*}[!t]
\centering
\begin{subfigure}{.3\textwidth}
  \centering
  \includegraphics[width=\textwidth]{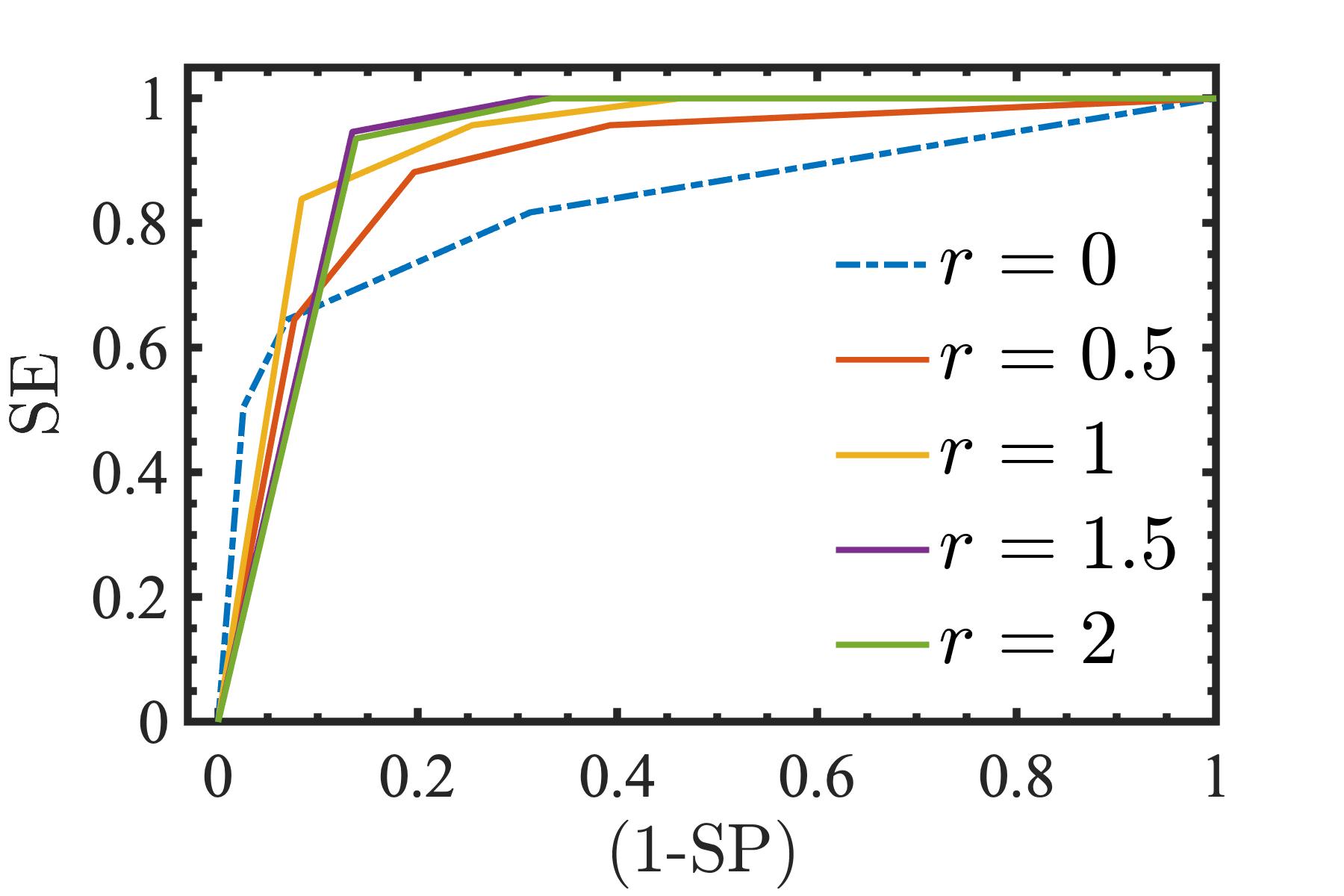}
  \caption{}
    \label{ch6_fig6}
\end{subfigure}%
\hfill
\begin{subfigure}{.3\textwidth}
    \centering
    \includegraphics[width=\textwidth]{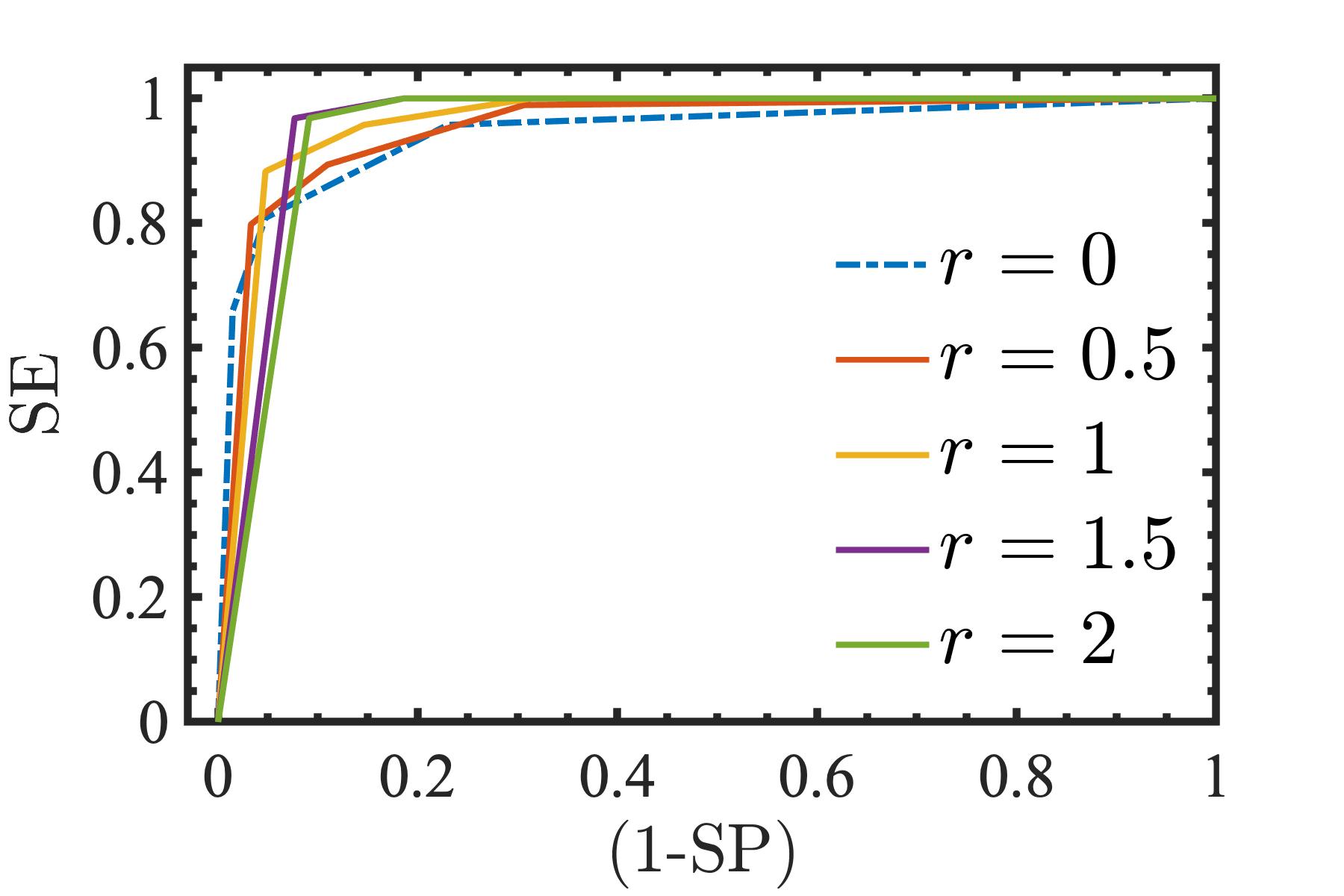}
    \caption{}
    \label{ch6_fig7}
\end{subfigure}
\hfill
\begin{subfigure}{.3\textwidth}
    \centering
    \includegraphics[width=\textwidth]{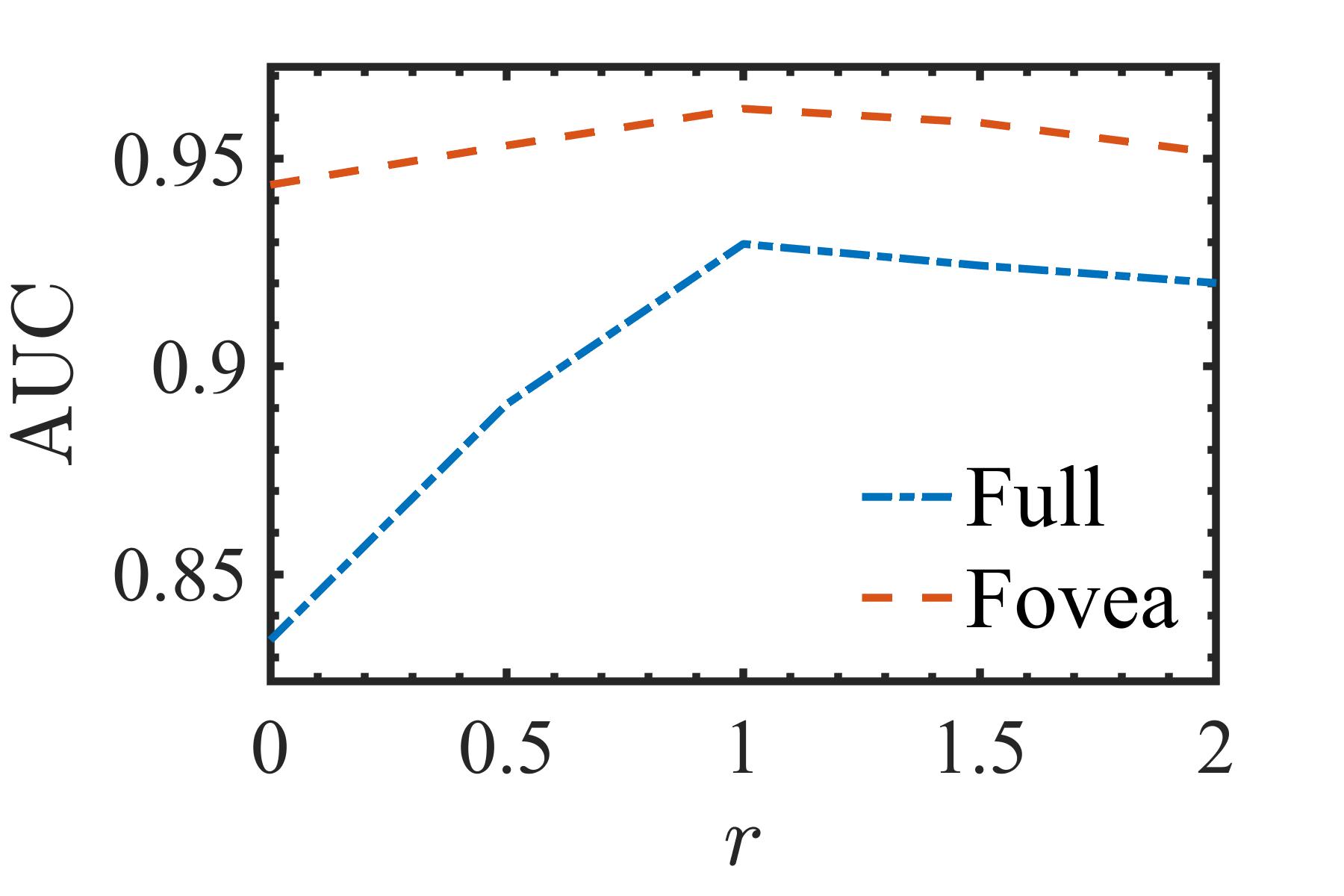}
    \caption{}
    \label{ch6_fig4}
\end{subfigure}
\caption{Effects of over-sampling ratio, $r$. (a) ROC curves obtained for full-images (b) ROC curves obtained for fovea-images (c) Variation in AUC.}
\label{f:compFullFovea}
\end{figure*}

\begin{table}[!b]
  \centering
  \caption{Effects of Class Balancing}
  \label{t:OS}%
  \begin{adjustbox}{max width=0.5\textwidth}
    \begin{tabular}{cccccc}
    \toprule
    \multicolumn{2}{c}{\textbf{Dataset}} & \makecell{\textbf{Sensitivity} \\ \textbf{(SE)}} & \makecell{\textbf{Specificity} \\ \textbf{(SP)}} & \makecell{\textbf{Overall}\\ \textbf{Accuracy (\boldmath$\eta$)}} & \textbf{AUC} \\
    \midrule
    \multirow{2}{*}{\makecell{Full\\ Image}} & \makecell{Original Data\\ (r=0)} & 0.6452 & 0.9309 & 0.8587 & 0.8365 \\
    \cmidrule{2-6}          &  \makecell{Oversampling\\ (r=1)} & 0.8387 & 0.9164 & 0.8967 & 0.9295 \\
    \midrule
    \multirow{2}[4]{*}{Fovea} & \makecell{Original Data\\ (r=0)} & 0.8085 & 0.9526 & 0.9158 & 0.9435 \\
    \cmidrule{2-6}          & \makecell{Oversampling\\ (r=1)} & 0.883 & 0.9526 & 0.9348 & 0.9621 \\
    \bottomrule
    \end{tabular}%
 \end{adjustbox}
\end{table}%

Fig. \ref{ch6_fig6} and \ref{ch6_fig7} show the ROC curves obtained with different values of $r$. The corresponding change in AUC is depicted in Fig.~\ref{ch6_fig4} which clearly show the positive effect of class balancing irrespective of the value of $r$. Further, it is interesting to see that there exist a non-monotonic relationship between $r$ and AUC. This is expected as while the over-sampling may improve sensitivity of the classifier, it may involve a trade-off in the specificity. It is easy to follow that this trade-off is optimized for $r=1$ for both \textit{full} and \textit{fovea} images as shown in Fig.~\ref{ch6_fig4}. Hence, for the rest of this study, the over-sampling ratio is fixed to `1', \textit{i.e.}, $r=1$.

Table \ref{t:OS} shows the classification performance over the test images with the original data (\textit{without over-sampling}, $r=0$) and with over-sampling ($r=1$). These results clearly indicate that the oversampling of the positive/minority class improves the classifier performance. In particular, for \textit{full} images, a significant improvement in AUC is obtained with over-sampling; from 0.8365 (with $r=0$) to 0.9295 (with $r=1$). 

It is interesting to see that, with the \textit{fovea} images, there is an improvement in the sensitivity without any compromise in the specificity. In contrast, for the \textit{full} images, while there is a significant improvement in the sensitivity (\textit{from 0.6452 with $r=0$ to 0.8387 with $r=1$}), it involves a slight trade-off in the specificity (\textit{from 0.9309 with $r=0$ to 0.9164 with $r=1$}). Nevertheless, such a trade-off is often acceptable in medical imaging tasks where the focus is on the positive screening accuracy (sensitivity). 

\subsection{Role of Feature Selection}
\label{res:fet_sel}

The objective of this part of the study is to investigate the effects of feature selection. For this purpose, a meta-heuristic feature selection approach (see Section~\ref{s:fet_sel}) is followed to identify the subset of optimal features, $\mathcal{X}^{\star}$, given by~(\ref{eq:fsp}), for both \textit{full} and \textit{fovea} images. A total of 40 independent runs of both GA and BPSO are carried out following the procedure outlined in Algorithm~\ref{al:ovfs}. Further, these algorithms are compared on the basis of two key objectives of feature selection, \textit{i.e.}, the \textit{classification performance}, $J(\cdot)$ and the \textit{number of features} (denoted by $\xi$). 

The outcomes of the feature selection such as the mean, variance and the best values of $J(\cdot)$ and $\xi$ which are obtained over multiple independent runs of the search algorithms are shown in Table~\ref{t:FS1}. In addition, the following two metrics are defined for purpose of comparative evaluation:
\begin{align}
    PI (\%) & =\frac {J(\mathcal{X}_{prime})-\overline{J(\mathcal{X}^*)}}{J(\mathcal{X}_{prime})} \times 100\\
    \Xi (\%) & = \frac {n - \xi_{avg}}{n} \times 100
\end{align}
where, `$J(\mathcal{X}_{prime})$' denotes the criterion function with the full feature set, $\mathcal{X}_{prime}$; $\overline{J(\mathcal{X}^*)}$  and `$\xi_{avg}$' denote respectively the average of the criterion function and cardinality which are obtained over 40 runs of the search algorithm; `$n$' is the total number of features (\textit{i.e.}, $n=1000$). It is clear that a higher values of $PI (\%)$ and $\Xi (\%)$ imply a better search performance.

A significant reduction in the cardinality (\textit{number of features}) is obtained for both \textit{full} and \textit{fovea} images; the reduction in cardinality, $\Xi (\%)$, lie in the range of $45\%-53\%$, as seen in Table~\ref{t:FS1}. In addition, the positive high values of $PI(\%)$ metric imply that the reduced feature subsets yield a performance improvement over the original feature set, $\mathcal{X}_{prime}$. It is interesting to see that, compared to BPSO, GA could identify a better subset for both \textit{full} and \textit{fovea} images. This further suggests the need for an effective search algorithm.

\begin{figure}[!t]
\centering
\begin{subfigure}{.3\textwidth}
  \centering
  \includegraphics[width=\textwidth]{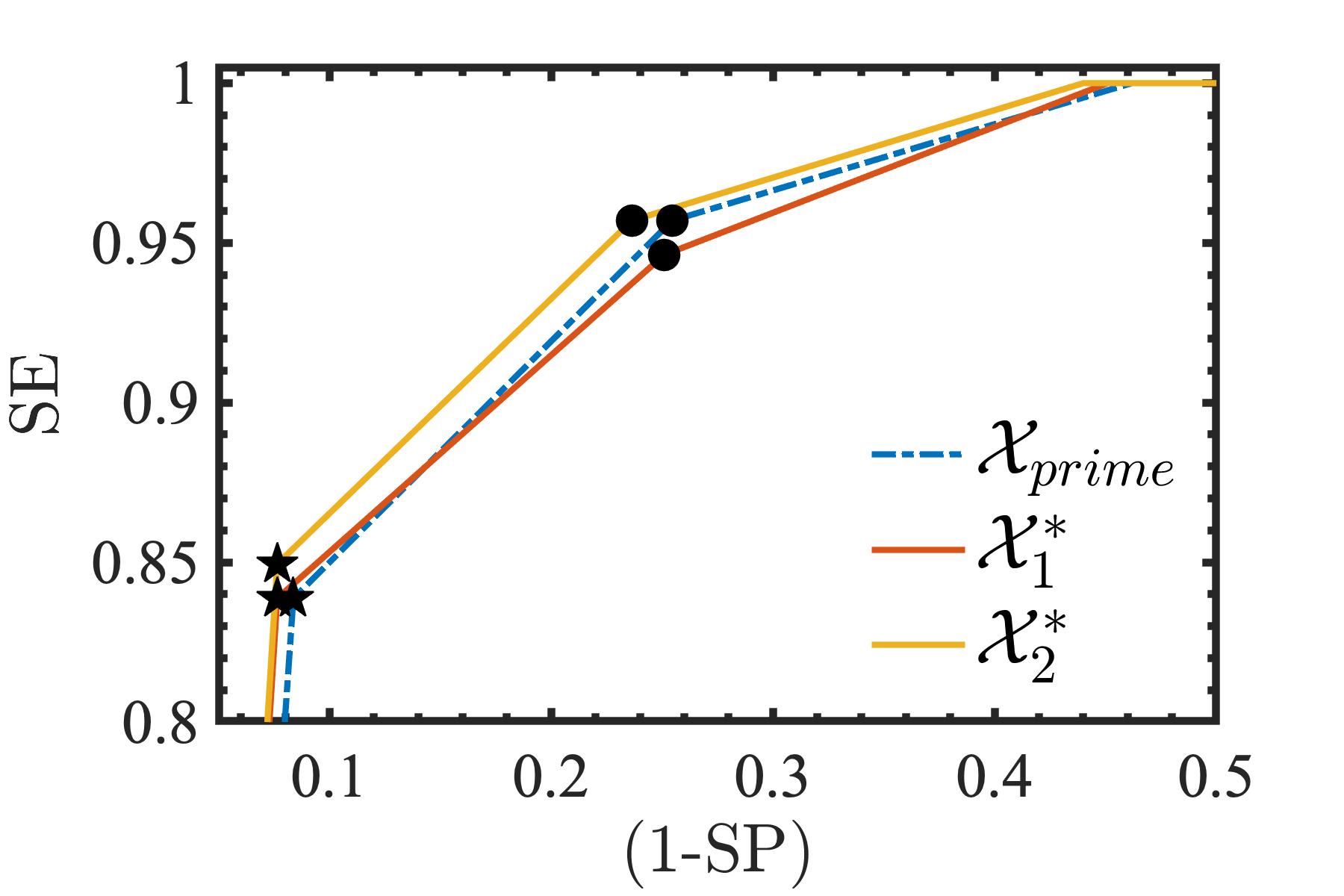}
  \caption{Full Image}
    \label{ch6_fig_FS1}
\end{subfigure}%
\hfill
\begin{subfigure}{.3\textwidth}
    \centering
    \includegraphics[width=\textwidth]{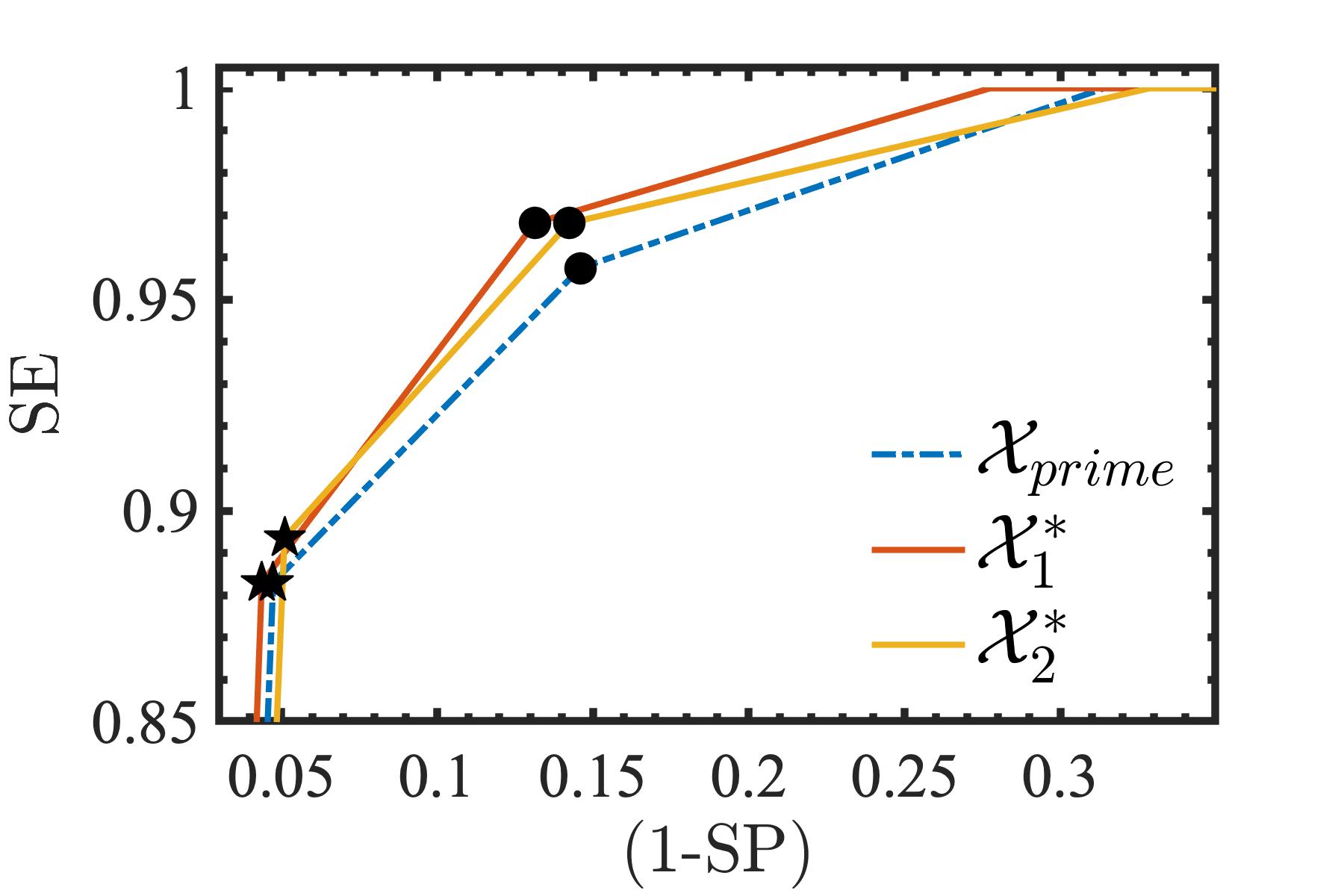}
    \caption{Fovea}
    \label{ch6_fig_FS2}
\end{subfigure}
\caption{The ROC curve obtained with testing images. $\mathcal{X}_{prime}$, $\mathcal{X}_{1}^\ast$ and $\mathcal{X}_{2}^\ast$ respectively denote the full feature set and the reduced feature subset identified by GA and BPSO. The Operating Point-A are marked with `$\star$' symbol on each ROC curve. Similarly, the Operating Point-B are marked with `$\bullet$'. }
\label{f:FSeffect}
\end{figure}

For further analysis, the `\textit{best}' subset identified over multiple runs of the algorithm is determined, as outlined in Line \ref{Ch6_line:3}-\ref{Ch6_line:4} (Algorithm~\ref{al:ovfs}). The \textit{best} subset identified by GA and BPSO are denoted here by `$\mathcal{X}_{1}^\ast$' and `$\mathcal{X}_{2}^\ast$', respectively. These subsets are further validated considering a set of 368 test images (see Section~\ref{s:database}). The ROC curve obtained for this test are shown in Fig. \ref{ch6_fig_FS1} (\textit{full}) and Fig. \ref{ch6_fig_FS2} (\textit{fovea}). The corresponding AUC values are listed in Table~\ref{t:FS2}. It is interesting to see that with significantly fewer features (\textit{approximately} $50\%$ of $n$), the reduced feature subsets, $\mathcal{X}_{1}^\ast$ and $\mathcal{X}_{2}^\ast$, could move the ROC curve towards the \textit{ideal} operating point located at the upper-left corner (see Fig~\ref{f:FSeffect}). This is also reflected by a slight increase in AUC values shown in Table~\ref{t:FS2}.

\begin{table}[!t]
  \centering
  \small
  \caption{Comparative Evaluation of Feature Selection Algorithms$^\dagger$} 
  \label{t:FS1}%
  \begin{adjustbox}{max width=0.42\textwidth}
  \begin{threeparttable}
    \begin{tabular}{cccc|cc}
    \toprule
    \multicolumn{2}{c}{\multirow{2}[4]{*}{\textbf{Results}}} & \multicolumn{2}{c|}{\textbf{Full Image}} & \multicolumn{2}{c}{\textbf{Fovea}} \\
    \cmidrule{3-6}    \multicolumn{2}{c}{} & \textbf{GA} & \textbf{BPSO} & \textbf{GA} & \textbf{BPSO} \\[0.5ex]
    \midrule
    \multirow{4}{*}{$J(\cdot)$} & Mean  & 0.0323 & 0.0375 & 0.0300 & 0.0339 \\[0.5ex]
          & SD    & 1.05E-03 & 1.31E-03 & 7.18E-04 & 1.01E-03 \\[0.5ex]
          & PI(\%) & 42.9  & 33.7 & 31.1  & 22.2 \\[0.5ex]
          & Best  & 0.0304 & 0.0347 & 0.0286 & 0.0321 \\[0.5ex]
    \midrule
    \multirow{4}{*}{$\xi(\cdot)$} & Mean  & 483.3 & 542.9 & 477.3 & 538.2 \\[0.5ex]
          & SD    & 17.43 & 17.55 & 15.64 & 20.20 \\[0.5ex]
          & Xi (\%) & 51.7  & 45.7 & 52.3  & 46.2 \\[0.5ex]
          & Best  & 462   & 558   & 467   & 502 \\
    \bottomrule
    \end{tabular}%
    \begin{tablenotes}
      \scriptsize
       \item $^\dagger$  Criterion function with full feature set: Full Images - $J(\mathcal{X}_{prime})=0.0566$, Fovea - $J(\mathcal{X}_{prime})=0.0436$
    \end{tablenotes}
    \end{threeparttable}
    \end{adjustbox}
\end{table}%
\begin{table}[!t]
  \centering
  \small
  \caption{Performance of Reduced Subsets on Test Images}
  \label{t:FS2}%
   \begin{adjustbox}{max width=0.38\textwidth}
    \begin{tabular}{cccc}
    \toprule
    \multirow{2}[4]{*}{\textbf{Dataset}} & \multicolumn{3}{c}{\textbf{AUC}} \\
    \cmidrule{2-4}          & \makecell{\textbf{Full} \\ \textbf{Set (\boldmath$\mathcal{X}_{prime}$)}} & \makecell{\textbf{GA} \textbf{(\boldmath$\mathcal{X}_1^\ast$)}} & \makecell{\textbf{BPSO} \textbf{(\boldmath$\mathcal{X}_2^\ast$)}} \\
    \midrule
    Full Image & 0.9295 & 0.9315 & 0.9362 \\[0.5ex]
    Fovea & 0.9621 & 0.9667 & 0.9625 \\
    \bottomrule
    \end{tabular}%
    \end{adjustbox}
\end{table}%
\subsection{Role of Fovea Segmentation}
\label{res:fov_seg}

In this part of the study, we investigate the effects of \textit{fovea} segmentation. For this purpose, the region within 1 DD around the fovea is segmented following the procedure outlined in~\cite{Chalakkal2018a}. Such segmentation is likely to improve the classification performance; as it limits the field of view to the region specific to CSME. This has been the motivation behind the fovea segmentation. The comparative analysis of the results with \textit{full} and \textit{fovea} images supports this notion. For instance, the results in Fig.~\ref{f:compFullFovea} show a marked improvement in the ROC curves with \textit{fovea} images. In particular, it is clear that, in comparison to \textit{full} images (Fig.~\ref{ch6_fig6}), this segmentation induces a positive shift in ROC curves towards the \textit{ideal} operating point (Fig.~\ref{ch6_fig7}). This shift in ROC translates into improved values of AUC with \textit{fovea} images as seen in Fig.~\ref{ch6_fig4}.

Similar results are also observed after feature selection. For instance, the values of AUC obtained with the reduced feature subsets ($\mathcal{X}_1^\ast$ and $\mathcal{X}_2^\ast$) lie in the range of $[0.93-0.94]$ for \textit{full} images; whereas these lie in the range of $[0.96-0.97]$ for \textit{fovea} images (see Table~\ref{t:FS2}).


\subsection{Selection of the ROC Operating Point}
\label{res:roc_opera}

The final part of this study focuses on the issues associated with the selection of ROC operating point and its implications. It is easy to follow that the selection of a particular ROC operating point essentially involves a trade-off in the cost associated with misdiagnosis, \textit{i.e.}, False Negative (FN) and False Positive (FP). Hence, this decision is primarily driven by the disease \textit{prevalence} in population. For instance, the costs associated with FN are relatively higher for lower values of prevalence. In such a scenario, the ROC operating point can be moved towards the right. 
In New Zealand, it is estimated that $6\%$ of the population is affected by diabetes, and $25\%$ of diabetics are diagnosed with CSME~\cite{Chan_2018}. Consequently, the prevalence of CSME is determined to be $1.5\%$. Hence, it is appropriate to move the operating point towards right to reduce the costs associated with FN. To highlight the effects of such selection, following two operating points are being considered:
\begin{itemize}
    \item  \textit{Operating Point-A}: This point represents the best overall cost compromise for both FP and FN. 
    \item \textit{Operating Point-B}: This point favors FN over FP. 
\end{itemize}

The Operating Point-A on ROC curves associated with $\mathcal{X}_{prime}$, $\mathcal{X}_1^\ast$ and $\mathcal{X}_2^\ast$ are respectively denoted by $A_1$, $A_2$ and $A_3$. Similar notations are also followed for the Operating Point-B. These points are shown in Fig. \ref{ch6_fig_FS1} (\textit{full}) and Fig. \ref{ch6_fig_FS2} (\textit{fovea}). 

The performance metrics corresponding to distinct operating points are given in Table \ref{t:FS3}. It is easy to follow that any departure from the optimal operating point (\textit{i.e.}, `A') essentially involves a trade-off in the overall accuracy. In particular, the shift to the Operating Point B will lead to the reduction in FN. While this will lead to a higher sensitivity, it comes with a trade-off in specificity. The results in Table \ref{t:FS3} clearly highlight such trade-off.

To summarize, the selection of the operating point is heavily dependent on the prevailing scenarios in a population. The operating points similar to `B' are often suitable when the disease prevalence is low and therefore a higher sensitivity is desirable.

\begin{table}[!t]
  \centering \small
  \caption{Implications of ROC Operating Point Selection}
  \label{t:FS3}%
  \begin{adjustbox}{max width=0.42\textwidth}
    \begin{tabular}{cccccc}

    \toprule
    \textbf{Images} & \textbf{Subset} & \makecell{\textbf{ROC}\\ \textbf{Point}} & \textbf{SE} & \textbf{SP} & \textbf{Accuracy} \\[0.5ex]
    \midrule
    \multirow{6}{*}{\makecell{Full\\ Image}} & \multirow{2}{*}{\makecell{Full\\[0.5ex] ($\mathcal{X}_{prime}$)}} & A1    & 0.8387 & 0.9164 & 0.8965 \\[0.5ex]
          &       & B1    & 0.9570 & 0.7455 & 0.7995 \\[0.5ex]
    \cmidrule{2-6}          & \multirow{2}[2]{*}{GA ($\mathcal{X}_1^\ast$)} & A2    & 0.8387 & 0.9236 & 0.9019 \\[0.5ex]
          &       & B2    & 0.9462 & 0.7491 & 0.7994 \\[0.5ex]
    \cmidrule{2-6}          & \multirow{2}[2]{*}{BPSO ($\mathcal{X}_2^\ast$)} & A3    & 0.8495 & 0.9236 & 0.9047 \\[0.5ex]
          &       & B3    & 0.9570 & 0.7636 & 0.8130 \\[0.5ex]
    \midrule
    \multirow{6}[6]{*}{Fovea} & \multirow{2}{*}{\makecell{Full\\ ($\mathcal{X}_{prime}$)}} & A1    & 0.8830 & 0.9526 & 0.9348 \\[0.5ex]
          &       & B1    & 0.9574 & 0.8540 & 0.8804 \\[0.5ex]
    \cmidrule{2-6}          & \multirow{2}[2]{*}{GA ($\mathcal{X}_1^\ast$)} & A2    & 0.8830 & 0.9562 & 0.9375 \\[0.5ex]
          &       & B2    & 0.9681 & 0.8686 & 0.8940 \\[0.5ex]
    \cmidrule{2-6}          & \multirow{2}[2]{*}{BPSO ($\mathcal{X}_2^\ast$)} & A3    & 0.8936 & 0.9489 & 0.9348 \\[0.5ex]
          &       & B3    & 0.9681 & 0.8577 & 0.8859 \\
    \bottomrule

    \end{tabular}%
    \end{adjustbox}
\end{table}%

\section{Conclusions}
\label{s:conclusion}

A new automated CSME screening has been proposed and its performance is thoroughly evaluated. The proposed approach alleviates the issues of \textit{exudate segmentation} and \textit{class imbalance}. The exudate segmentation is replaced by a combination of DCNN and a meta-heuristic feature selection approach. The problems associated with imbalance data sets have been overcome by minority class oversampling. The results of comprehensive investigation show that by judicious integration of few data-processing steps such as class balancing and feature selection, it is possible to accomplish the screening task with any basic induction algorithm, such as k-NN. Although, the focus of this study has been on CSME screening, the proposed approach could easily be tailored for other medical imaging tasks.

\bibliographystyle{IEEEtran}


\end{document}